\pdfoutput=1

\documentclass[11pt]{article}
\usepackage{jheppub}

\usepackage{amsmath}
\usepackage{amssymb}
\usepackage{graphicx,color,slashed,mathtools}
\usepackage{bbm} 
\usepackage{ifpdf}

\newcommand{\be}{\begin{equation}}
\newcommand{\ee}{\end{equation}}
\newcommand{\ba}{\begin{eqnarray}}
\newcommand{\ea}{\end{eqnarray}}

\newcommand{\lp}{\left(}
\newcommand{\rp}{\right)}

\newcommand{\N}{\mathcal{N}}

\def\rmi{{\rm i}}

\newcommand{\Db}{\overline{D6}}

\newcommand{\V}{\mathcal{V}}

\title{Scaling limits of dS vacua and the swampland}

\author{Andreas  Banlaki,}
\author{Abhishek Chowdhury,}
\author{Christoph Roupec,}
\author{Timm Wrase\\}

\affiliation{Institute for Theoretical Physics, TU Wien,\\
Wiedner Hauptstrasse 8-10/136, A-1040 Vienna, Austria\\}

\emailAdd{banlaki@hep.itp.tuwien.ac.at} % Is there a better email address?
\emailAdd{abhishek.chowdhury@tuwien.ac.at}
\emailAdd{christoph.roupec@tuwien.ac.at}
\emailAdd{timm.wrase@tuwien.ac.at}

\notoc

\abstract{ 
\noindent We discuss the properties of massive type IIA flux compactifications. In particular, we investigate in which case one can obtain dS vacua at large volume and small coupling. We support a general discussion of scaling symmetries with the analysis of a concrete example. We find that the large volume and weak coupling limit requires a large number of O6-planes. Since these are bound for any given compactification space one cannot get arbitrarily good control over $\alpha'$ and string loop corrections.

}

\begin{document}

\maketitle

\newpage

\section{Introduction}\label{sec:introduction}
Recently there has been a revived interest in dS vacua in string theory due to the so called dS swampland conjecture \cite{Obied:2018sgi}, that says that there are no dS vacua in string theory.\footnote{This renewed interest in the topic has led to a surge of new papers, see for example \cite{Agrawal:2018own, Andriot:2018wzk, Dvali:2018fqu, Achucarro:2018vey, Garg:2018reu, Lehners:2018vgi, Kehagias:2018uem, Dias:2018ngv, Denef:2018etk, Colgain:2018wgk, Paban:2018ole, Roupec:2018mbn, Andriot:2018ept, Matsui:2018bsy, Ben-Dayan:2018mhe, Heisenberg:2018yae, Damian:2018tlf, Conlon:2018eyr, Kinney:2018nny, Dasgupta:2018rtp, Cicoli:2018kdo, Kachru:2018aqn, Akrami:2018ylq, Murayama:2018lie, Marsh:2018kub, Brahma:2018hrd, Choi:2018rze, Danielsson:2018qpa, DAmico:2018mnx, Han:2018yrk, Moritz:2018ani, Halverson:2018cio, Ellis:2018xdr, Anguelova:2018vyr, Lin:2018kjm, Hamaguchi:2018vtv, Ashoorioon:2018sqb, Ooguri:2018wrx, Wang:2018kly, Fukuda:2018haz, Hebecker:2018vxz, Gautason:2018gln, Olguin-Tejo:2018pfq, Garg:2018zdg, Dvali:2018jhn, Park:2018fuj, Agrawal:2018rcg, Heckman:2018mxl, Yi:2018dhl, Chiang:2018lqx, Cheong:2018udx, Ibe:2018ffn, Blanco-Pillado:2018xyn}.} While currently quintessence models with for example simple exponential potentials provide an observationally viable alternative to dS vacua \cite{Agrawal:2018own, Akrami:2018ylq}, it is unclear whether such models are easier to construct in string theory than dS vacua \cite{Cicoli:2018kdo}. Given that current and future experiments are testing a very interesting parameter space, namely quintessence models with $V(\phi) \propto e^{c \phi}$ for $c\lesssim 1$, it is of uttermost importance to better understand quintessence in string theory as well as the dS swampland conjecture.

In this paper we will not discuss quintessence but we rather focus on the dS swampland conjecture and the status of classical dS vacua in string theory. Similar to most of the swampland conjectures it is difficult to actually prove the dS swampland conjecture without a non-perturbative understanding of string theory. So a simpler avenue for progress is to try to disprove the dS swampland conjecture.\footnote{Failure might then provide further evidence for the conjecture and/or it gives us a deeper understanding of dS space in general.} In order to scrutinize it let us quickly recap what it is based on: Firstly, there has been an on-going debate in the community about the viability of the existing constructions of dS vacua in string theory, in particular the KKLT \cite{Kachru:2003aw} and LVS \cite{Balasubramanian:2005zx} scenarios, see \cite{Danielsson:2018ztv} for a recent review. Secondly, there are no dS constructions, even phenomenological uninteresting ones in higher or lower dimensions, that are comparably as simple as AdS or Minkowski vacua (see \cite{VanRiet:2011yc} for related work). Thirdly, the authors of \cite{Obied:2018sgi} present a large number of explicit examples that satisfy the dS swampland conjecture.

The first two points that were listed in support of the conjecture seem to be the ones where immediate progress can be made. In particular, the status of KKLT is the content of active research right now \cite{Moritz:2017xto, Sethi:2017phn, Kachru:2018aqn, Kallosh:2018wme, Danielsson:2018qpa, Kallosh:2018psh}, so one can be hopeful that this ongoing work will ultimately lead to a consensus in the community on the status of the KKLT scenario. However, given the fact that the discussion about the consistency of the KKLT construction has started almost a decade ago \cite{Bena:2009xk}, it is worthwhile to simultaneously explore other dS constructions in string theory. In particular, the original dS swampland paper \cite{Obied:2018sgi} discusses fairly simple classical flux compactifications in support of the conjecture. These construction evade the no-go theorem of Maldacena-Nu\~nez \cite{Maldacena:2000mw} because they allow for O-planes. Nevertheless, there are often other no-go theorems, summarized in  \cite{Wrase:2010ew}, that generalize the original no-go theorem of \cite{Hertzberg:2007wc} (see \cite{Andriot:2016xvq, Andriot:2017jhf, Andriot:2018ept} for more recent results). It was however shown that some classes of such compactification evade all no-go theorems and allow for dS critical points  \cite{Caviezel:2008tf, Flauger:2008ad, Danielsson:2009ff, Caviezel:2009tu, Danielsson:2010bc, Andriot:2010ju, Petrini:2013ika, Roupec:2018mbn}. These existing solutions seem to be often at strong coupling or small volume and should receive substantial $\alpha'$ and/or string loop corrections. While there have been recent breakthroughs that show that the obstinate tachyon \cite{Shiu:2011zt, Danielsson:2012et, Junghans:2016uvg, Junghans:2016abx} in these dS critical points can be removed by including anti-D6-branes \cite{Kallosh:2018nrk} or by adding KK monopoles \cite{Blaback:2018hdo}, it remains unclear whether one can find dS vacua that are at large volume and weak coupling, i.e. that are in a regime in which we can trust the supergravity approximation. We address this point in this paper by discussing scaling limits of the four dimensional scalar potential. 

{\bf Note added:} While we were finalizing this paper we were informed that a related paper is about to appear \cite{Junghans:2018gdb}. While the motivation for the two papers is different, they both show that dS solutions at parametrically good control cannot exist in massive type IIA flux compactifications with O6-planes sources, although this seems to be possible for supersymmetric AdS vacua \cite{DeWolfe:2005uu}.

\section{Scaling limits of AdS and dS vacua}
In this section we discuss the scaling limits of the scalar potential that arises in flux compactifications of massive type IIA string theory. While the full scalar potential is rather complicated, one can learn already a lot by restricting to a two dimensional slice in moduli space that is spanned by the string coupling $e^\phi$ and the internal volume $\V_6$. In particular, if we define the following scalars
\be
\rho = (\V_6)^\frac13\,, \qquad \quad \tau = e^{-\phi} \sqrt{\V_6}\,,
\ee
then the scalar potential takes the form
\be\label{eq:V}
V(\rho,\tau) = \frac{A_H}{\rho^3 \tau^2} + \sum_{p=0,2,4,6} \frac{A_{F_p}}{\rho^{p-3} \tau^4} - \frac{A_{sources}}{\tau^3}+\frac{A_{R_6}}{\rho \tau^2}\,.
\ee
Here all $A$'s are positive definite, except $A_{sources}$ and $A_{R_6}$. $A_H$ arises from integrating the NSNS field strength term, $|H|^2$, in the ten dimensional action over the internal space. Similarly, the $A_{F_p}$ correspond to the contributions from the RR-fluxes $F_p$. All of these terms depend on the flux quanta in a quadratic way and also on all other moduli that we are suppressing in this discussion. As sources we want to include O6-planes, D6-branes as well as anti-D6-branes. The term $A_{sources}$ arises from the DBI-terms integrated over the 3-cycles wrapped by these sources and therefore has the following dependence on the number of O6-planes and D6-branes: $A_{sources} \propto 2N_{O6}-N_{D6}-N_{\Db}$. The suppressed prefactor is positive since (anti-)D-branes have positive tension and therefore contribute positively to the scalar potential. Lastly, $A_{R_6}\propto -R_6$, arises from integrating the internal Ricci scalar over the internal space. It is positive, if the internal space is negatively curved and negative, if the internal space is positively curved.

\subsection{A Maldacena-Nu\~nez type no-go theorem}
Let us start out by reproducing a variant of the well-known no-go theorem against dS vacua in the absence of orientifold planes \cite{Maldacena:2000mw}. In our scalar potential above the absence of O6-planes, $N_{O6}=0$, is not really distinguishable from the more general case with $2N_{O6}-N_{D6}-N_{\Db} <0$. So we will be able to conclude that whenever the total number of D6-branes and anti-D6-branes is larger than twice the number O6-planes then no dS solutions can exist. To proof this we simply minimize $V$ with respect to $\tau$
\be
0=\partial_\tau V = -2 \frac{A_H}{\rho^3 \tau^3} -4 \sum_{p=0,2,4,6} \frac{A_{F_p}}{\rho^{p-3} \tau^5} -3 \frac{|A_{sources}|}{\tau^4}-2 \frac{A_{R_6}}{\rho \tau^3}\,.
\ee
For negatively curved manifolds with $A_{R_6} \propto -R_6 >0$ we see that there is no solution since all terms are negative definite. For positively curved spaces we can solve the above equation in terms of $A_{R_6}$
\be
\frac{A_{R_6}}{\rho \tau^2} = -\frac{A_H}{\rho^3 \tau^2} -2 \sum_{p=0,2,4,6} \frac{A_{F_p}}{\rho^{p-3} \tau^4} -\frac32 \frac{|A_{sources}|}{\tau^3}\,.
\ee
Plugging this back into $V$ we get
\be
V = - \sum_{p=0,2,4,6} \frac{A_{F_p}}{\rho^{p-3} \tau^4} -\frac12 \frac{|A_{sources}|}{\tau^3} <0\,.
\ee
So we see that the solution is necessarily AdS. We conclude that dS solutions can only exist for a bounded range of D6 and anti-D6-branes, $N_{D6}+N_{\Db} <2N_{O6}$.\footnote{Generically there are many 3-cycles that can be wrapped by O6-planes and (anti-)D6-branes. In that case we have $A_{sources} \propto \mathcal{Z}^K(2N_{O6}-N_{D6}-N_{\Db})_K$, where $\mathcal{Z}^K$ are complex structure moduli and $K$ runs over the number of 3-cycles. Here our result trivially generalizes to $\mathcal{Z}^K(2N_{O6}-N_{D6}-N_{\Db})_K>0$.}  Note that the number of O6-planes itself is given by the number of fixed points under the orientifold projection and therefore determined by the properties of the compactification space.

\subsection{Parametrically controlled supersymmetric AdS solutions}
Next we consider a very interesting but also somewhat surprising feature of supersymmetric AdS vacua of massive type IIA flux compactifications that was discovered in \cite{DeWolfe:2005uu}. There the authors showed that the $F_4$-flux, which is unconstrained by tadpole cancellation conditions, can be made arbitrarily large. In this limit in which all $F_4$-fluxes, which we denote by $f_4$, are large the authors find supersymmetric AdS solutions for flux compactifications on Calabi-Yau manifolds that have parametrically large volume $\V\propto (f_4)^{\frac32}$ and parametrically weak string coupling $e^{-\phi} \propto (f_4)^{\frac34}$. Furthermore, the four dimensional Hubble scale $H$ is parametrically smaller than the KK-scale $1/R$, $H R \propto (f_4)^{-\frac12}$, so that these solutions are truly four dimensional (contrary to for example Freund-Rubin type solutions \cite{Freund:1980xh}).\footnote{This scale separation between the four dimensional Hubble scale and the KK-scale is another important aspect of flux compactifications that was discussed for these setups in \cite{Gautason:2015tig}.} 

Let us study this interesting scaling limit in our simplified scalar potential. For Calabi-Yau manifolds we have $R_6=0$ and therefore $A_{R_6} = 0$. Furthermore, we take $A_{F_4} = a_{F_4}( f_4)^2$, $\rho =  (\V_6)^\frac13 = \tilde \rho (f_4)^{\frac12}$ and $\tau = e^{-\phi} \sqrt{\V_6} = \tilde \tau (f_4)^{\frac32}$ and find that the scalar potential becomes
\be
V(\rho,\tau) = \frac{1}{(f_4)^{\frac92}}\lp \frac{A_H}{\tilde\rho^3 \tilde\tau^2} +\frac{A_{F_0}}{\tilde\rho^{-3} \tilde\tau^4}+\frac{a_{F_4}}{\tilde\rho \tilde\tau^4} - \frac{A_{sources}}{\tilde\tau^3}\rp +\frac{1}{(f_4)^{\frac{11}{2}}} \frac{A_{F_2}}{\tilde\rho^{-1} \tilde\tau^4} +\frac{1}{(f_4)^{\frac{15}{2}}} \frac{A_{F_6}}{\tilde\rho^{3} \tilde\tau^4} \,.
\ee
Thus we see that unless we make the $F_2$- and/or the $F_6$-fluxes large, they will become irrelevant in this limit. However, as was shown in \cite{DeWolfe:2005uu} the $F_2$- and $F_6$-fluxes essentially only affect the values of the $B_2$ and $C_3$ axions at their minimum but they are not relevant for the actual stabilization of non-axionic moduli. So all the terms that are needed in order to stabilize the moduli survive in this scaling limit (as expected from the results in \cite{DeWolfe:2005uu}). Note that in this limit we can leave the number of O6-planes $N_{O6}$, as well as the $H$- and $F_0$-fluxes small. These are all three tied together via the tadpole condition, which reads for all 3-cycles $\Sigma_K$ in integer 3-homology 
\be
\sqrt{2} \int_{\Sigma_K} d (dC_1+F_0 B) = \sqrt{2} \int_{\Sigma_K} F_0 H =\left. (-2N_{O6}+N_{D6}+N_{\Db})\right|_\text{wrapped on $\Sigma_K$}\,.
\ee
So it seems that even for very small numbers of O6-planes, there exist supersymmetric AdS solutions with arbitrarily good control over all corrections. Such AdS$_4$ solutions should have CFT$_3$ duals with unusual properties, some of which have been studied in \cite{Aharony:2008wz}. We will not be able to add anything new to this but we would like to point out that in the presence of curvature the large $F_4$-flux limit is obstructed. Let us again look at the scalar potential in the limit $A_{F_4} = a_{F_4}( f_4)^2$, $\rho =  (\V_6)^\frac13 = \tilde \rho (f_4)^{\frac12}$ and $\tau = e^{-\phi} \sqrt{\V_6} = \tilde \tau (f_4)^{\frac32}$ but this time we include the curvature term $A_{R_6}$
\be
V(\rho,\tau) = \frac{1}{(f_4)^{\frac72}} \frac{A_{R_6}}{\tilde\rho \tilde\tau^2} +  \frac{1}{(f_4)^{\frac92}}\lp \frac{A_H}{\tilde\rho^3 \tilde\tau^2} +\frac{A_{F_0}}{\tilde\rho^{-3} \tilde\tau^4}+\frac{a_{F_4}}{\tilde\rho \tilde\tau^4} - \frac{A_{sources}}{\tilde\tau^3}\rp+\frac{1}{(f_4)^{\frac{11}{2}}} \frac{A_{F_2}}{\tilde\rho^{-1} \tilde\tau^4} +\frac{1}{(f_4)^{\frac{15}{2}}} \frac{A_{F_6}}{\tilde\rho^{3} \tilde\tau^4}  \,.
\ee
We see that the potential in the large $f_4$ limit has only a single leading term and therefore leads to a runaway for $\tilde \tau$ and $\tilde \rho$. This means no vacua can exist in this limit. The curvature contribution vanishes for large $\rho$ and $\tau$ since it scales like $A_{R_6}/(\rho \tau^2)$, however, in this particular limit of large $F_4$-flux it does not vanish as fast as the other terms and hence it actually dominates the potential. 

One might worry that localizing the O6-planes leads to backreaction effects that effectively introduce curvature. However, such a backreaction due to warping will change the identification of the moduli and the scalings in the effective potential, so that one cannot reach such a conclusion without a proper understanding of the warped effective field theory in these models. The general expectation is that corrections due to warping should become negligible for large volume and weak coupling so one would actually expect that these parametrically controlled AdS vacua cannot get destroyed by backreaction effects.\footnote{We like to thank Thomas Van Riet for discussion of this point.}

\subsection{Controlled dS vacua?}\label{ssec:dS}
Now we would like to study the existence of dS vacua in the controlled regime of large volume and weak coupling. First we recall some simple no-go theorems \cite{Haque:2008jz, Flauger:2008ad, Wrase:2010ew} that forbid dS vacua in these setting for vanishing mass parameter, i.e. $A_{F_0}=0$, and/or for vanishing or positive curvature $A_{R_6} \leq 0$. Let us look first at the case without mass parameter, which can in principle be lifted to M-theory. If we set $A_{F_0}=0$, then we find from equation \eqref{eq:V} that
\be\label{eq:bound}
(-\rho\partial_\rho-\tau \partial_\tau) V= 5 \frac{A_H}{\rho^3 \tau^2} + \sum_{p=2,4,6} (p+1)\frac{A_{F_p}}{\rho^{p-3} \tau^4} - 3\frac{A_{sources}}{\tau^3}+3\frac{A_{R_6}}{\rho \tau^2} \geq 3 V\,.
\ee
So we see that dS extrema cannot exist, since they would require $V>0$ and $\partial_\tau V=\partial_\rho V=0$. It is also instructive to calculate the lower bound on the first slow-roll parameter, which requires us to know the kinetic terms for $\rho$ and $\tau$. These can be obtained from dimensional reduction and are given by \cite{Hertzberg:2007wc}
\be
\mathcal{L} = \frac12 R - \frac34 \frac{\partial_\mu \rho \partial^\mu \rho}{\rho^2} -  \frac{\partial_\mu \tau \partial^\mu \tau}{\tau^2}+\ldots = \frac12 R - \frac12 \partial_\mu \hat\rho \partial^\mu \hat \rho - \frac12 \partial_\mu \hat\tau \partial^\mu \hat\tau+\ldots \,,
\ee
where the dots correspond to all the other scalar fields and the scalar potential. Thus we find
\be
\epsilon = \frac12 \frac{(\partial_{\hat\rho} V)^2 +(\partial_{\hat\tau} V)^2 +\ldots }{V^2} \geq  \frac13 \lp\frac{\rho \partial_{\rho} V}{V}\rp^2+\frac14\lp\frac{\tau \partial_{\tau} V}{V}\rp^2\,.
\ee
Using equation \eqref{eq:bound} we then find
\be
\epsilon \geq  \frac13 \lp3+\frac{\tau \partial_{\tau} V}{V}\rp^2+\frac14\lp\frac{\tau \partial_{\tau} V}{V}\rp^2\,.
\ee
Minimizing the right hand side with respect to $\frac{\tau \partial_{\tau} V}{V}$, one finds the bound $\epsilon \geq \frac{9}{7}$.

Similarly, for the case with $A_{R_6} \leq 0$, one finds that $(-\rho\partial_\rho-3\tau \partial_\tau) V\geq 9 V$, which after some algebra implies that $\epsilon \geq \frac{27}{13}$ \cite{Hertzberg:2007wc}. Thus, we can never find dS vacua in these setups unless we have a mass parameter and a compactification space with negative curvature.

Now we want to address the important question, when can we find solutions at large volume and weak coupling, i.e. at $\rho,\tau \gg 1$? We have seen in the previous subsection that the large $F_4$-flux limit that was present for AdS vacua obtained from compactifications on Calabi-Yau manifolds is obstructed in the presence of curvature. However, this does not mean that there cannot be another limit in which one can go to large volume and weak coupling.

We have seen above that dS vacua require $A_{sources}>0$, $A_{F_0}\neq 0$ and $A_{R_6}>0$. Let us make the two moduli simultaneously large by scaling them as follows $\rho \propto \lambda^{c_\rho}$ and $\tau \propto \lambda^{c_\tau}$ for $\lambda \rightarrow \infty$. If we keep $A_{sources}$, $A_{F_0}$ and $A_{R_6}$ fixed in this limit, then we find that the corresponding three terms in the scalar potential never scale the same way, since this would require the following conditions $3c_\tau=4c_\tau-3c_\rho=2c_\tau+c_\rho$ to have a non-trivial solution for $c_\tau$ and $c_\rho$. However, the only solution is the trivial one, $c_\tau=c_\rho=0$, in which case the volume and dilaton do not scale. Thus we find that there are no dS vacua at parametrically large volume and weak coupling, if we keep $A_{sources}$, $A_{F_0}$ and $A_{R_6}$ fixed. This is because, per the conditions above, all three terms needed to stay relevant in the large volume and weak coupling limit, which is impossible.

Now we want to ask which fluxes and/or sources we have to change in order to find controlled dS vacua. For that purpose we recall first that all terms in the scalar potential are quadratic in the flux quanta \cite{Herraez:2018vae}, with the curvature term being quadratic in the so called metric fluxes $\omega$. The only exception here is $A_{sources} \propto 2N_{O6}-N_{D6}-N_{\Db}$ which is linear in the number of O6-planes and (anti-)D6-branes. However, using the tadpole condition that for spaces with metric fluxes $\omega$ has the schematic form $\sqrt{2}\int \lp \omega\cdot F_2 +F_0 H \rp = -2N_{O6}+N_{D6}+N_{\Db}$, one can in principle replace this term proportional to the right-hand-side of the tadpole equation, with a term that is quadratic in the fluxes, i.e. the left-hand side of the tadpole equation. For us it does not matter whether we do this replacement but we notice that due to the tadpole $-2N_{O6}+N_{D6}+N_{\Db}$ has to scale like $\int \lp \omega\cdot F_2 +F_0 H \rp$. This means that the above result in the previous paragraph is actually generic: $-2N_{O6}+N_{D6}+N_{\Db}$ has to be negative and $N_{O6}$ is fixed by the particular orientifold projection to a finite number that cannot be scaled to become arbitrarily large. The tadpole condition then generically fixes $F_0$ and $\omega$ to be of the same order, which means that $A_{sources}$, $A_{F_0}$ and $A_{R_6}$ are generically fixed and not too different.

There is one loop-hole to the above generic statement. It is in principle possible to have $\sqrt{2}\int \lp \omega\cdot F_2 +F_0 H \rp= -2N_{O6}+N_{D6}+N_{\Db}$ smallish but both terms $\int \omega\cdot F_2$ and $\int F_0 H$ can become large, if they have opposite signs and cancel almost exactly. Let us study this limit by also allowing the scaling $A_{H} \propto \lambda^{c_H}$, $A_{F_p} \propto \lambda^{c_{F_p}}$ and $A_{R_6} \propto \lambda^{c_{R_6}}$. The requirement that the three terms in the scalar potential involving $A_{sources}$, $A_{R_6}$ and $A_{F_0}$ scale in the same way amounts to
\be
-3c_\tau = c_{R_6}-c_\rho-2c_\tau \qquad \text{and} \qquad -3c_\tau =c_{F_0} + 3 c_\rho -4c_\tau\,.
\ee
Solving the first equation, we find
\be
c_\tau =c_\rho - c_{R_6}\,.
\ee
Plugging the above in the second equation we find
\be
2 c_\rho = - c_{F_0}-c_{R_6}\,.
\ee
This is actually a contradiction. The reason for this is that all the terms are quadratic in the fluxes, which we can make large. However, all fluxes are bounded from below by turning on a single flux quanta, so we cannot make them small. So if $c_{F_0}$ and $c_{R_6}$ are positive then we find that we would need $c_\rho$ to be negative. This then would imply that the volume $\V = \rho^3 \propto \lambda^{3 c_\rho} = \lambda^{-3 |c_\rho|}$ shrinks in the $\lambda \rightarrow \infty$ limit.

So in summary, we have shown that no dS vacua with parametric control over $\alpha'$ and string loop corrections exist in this particular setup. However, this does not mean that one cannot find dS vacua with somewhat large volume and small string coupling. To achieve this one would need to have some freedom in establishing small hierarchies between the different $A$'s. This seems not really possible in models with a single O6-plane on each 3-cycles, like the ones studied in \cite{Danielsson:2011au},\footnote{\label{fn:NO6} Let us recall the proper counting for $T^6/\mathbb{Z}_2\times\mathbb{Z}_2$ and an O6-plane orientifold projection that reverses the signs of three circle directions. One expects two fixed point per circle, i.e. $2^3=8$ O6-planes in the covering space. In the quotient space we have to take into account that we have orbifolded by $\mathbb{Z}_2\times\mathbb{Z}_2$ and done another $\mathbb{Z}_2$ orientifold involution so we have to divide by $2^3$ and find $N_{O6}=1$.} since then usually the $H$-flux and $F_0$-flux are fixed to be one or two as well. However, for more complicated compactifications the number of fixed points under the orientifold projection can certainly be much larger and we would like to determine how this should allow us to find dS vacua at large volume and weak coupling. 

Above we kept $A_{sources}$ fixed, which led us to an inconsistency, if we want to have the same scaling for the terms of $A_{F_0}$, $A_{sources}$ and $A_{R_6}$ in \eqref{eq:V} and also wanted to satisfy the tadpole condition. Now we will also allow the source term $A_{sources} \propto \lambda^{c_s}$ to scale, corresponding to allowing for a large number $N_{O6}$ of O6-planes. Plugging this into the scalar potential \eqref{eq:V} we then find that $A_{F_0}$, $A_{sources}$ and $A_{R_6}$ scale in the same way with $\lambda$, if 
\be 
-3 c_\rho + 4 c_\tau -c_{F_0} = 3 c_\tau - c_s = c_\rho + 2 c_\tau - c_{R_6}.
\ee
From the tadpole condition$\sqrt{2}\int (\omega \cdot F_2 + F_0 \cdot H) = -2 N_{O6} + N_{D6} + N_{\Db}$ we find that generically we have to satisfy
\be 
\frac{1}{2} \left( c_{R_6} + c_{F_2} \right) = \frac{1}{2} \left( c_{F_0} + c_{H} \right) = c_s,
\ee
where the factors of $\frac{1}{2}$ arise because the flux contributions in the scalar potential are quadratic in the fluxes, while the fluxes appear linear in the tadpole condition and on the right-hand-side only $N_{O6} \propto A_{sources}$ scales. Considering all these conditions we find the solution
\ba \label{eq:scalsol}
c_{\rho} &=& c_s - \frac{1}{2} \left( c_{F_0} + c_{R_6} \right)\,,\cr
c_{\tau} &=& 2 c_s - \frac{1}{2} \left( c_{F_0} + 3 c_{R_6}\right)\,,\cr
c_{F_2} &=& 2 c_s - c_{R_6}\,,\cr
c_H &=& 2 c_s - c_{F_0}\,.
\ea
It is now interesting to check how the volume and the string coupling scale with the number of O6-planes. We have that $\rho = \V_6^{\frac{1}{3}}$ and $\tau  = e^{-\phi} \sqrt{\V_6}$, using our solution \eqref{eq:scalsol} this translates to
\be 
\V_6 \propto \lambda^{\frac{3}{2} \left[ 2 c_s - ( c_{F_0} + c_{R_6} )\right]} \qquad \text{and} \qquad e^{-\phi} \propto \lambda^{\frac{1}{4} \left[  2 c_s + c_{F_0} - 3 c_{R_6}  \right]}.
\ee
If we want to have a trustworthy solution, we require large volume and small coupling. A sufficient constraint for this is
\be \label{eq:const}
c_s > \frac{3}{2} c_{R_6} + \frac{1}{2} c_{F_0}.
\ee 
Let us look at one special case where we set $c_{F_0} = c_{R_6} = 0$. The volume and string coupling then scale with the number of O6-planes, $N_{O6}$, as follows
\be\label{eq:largeNO6}
\V_6 \propto N_{O6}^{3} \qquad \text{and} \qquad e^{-\phi} \propto \sqrt{N_{O6}}\,.
\ee
Thus, we see that a large number of orientifold plane can lead to controlled compactifications with large volume and small string coupling. While for type IIB and F-theory compactification on Calabi-Yau manifolds it is very well understood what values one can expect for the number of O3-plane charges, we are not aware of similar results for type IIA compactifications on SU(3)-structure manifolds. It would be very interesting to understand this better and ideally find explicit models with large numbers of O6-planes or show on general grounds what is the upper bound on $N_{O6}$.

Note, that our discussion above neglects all order one coefficients that might or might not help in obtaining large volume and weak coupling solutions. Either way it seems clear from our results that a large number of O6-planes would be desired in order to find trustworthy dS solutions. In the next section we will study an explicit example in which we have analytic control over the solutions and we will reproduce the above scalings with the number of O6-plane including all order one factors. Before we do that we would like to recall which other ingredients could be added in these setups.

\subsection{Other ingredients}
There are many other ingredients one can include in flux compactifications of massive type IIA. Similar to the fluxes, where the absence of one flux seems very special, one expects that the most generic compactification is the one with all possible ingredients included. However, each ingredient we add makes the compactification more complex, leads to more complicated backreaction effects and introduces potentially new degrees of freedoms (like the ones localized on branes). Therefore, the approach above should be understood as the search for a minimalistic compactification that could give rise to dS vacua.
 
In particular, natural ingredients to add might be NS5-branes or KK-monopoles. After all, we are including the corresponding fluxes so it seems natural to allow for the sources as well. Here the scaling for NS5-branes, NSO5-plane, KK-monopoles and corresponding KKO-planes is (see for example \cite{Silverstein:2007ac} for the explicit expressions)
\be
V_{NS5} = \frac{A_{NS5}}{\rho^2 \tau^2}\,, \quad V_{NSO5} = -\frac{A_{NSO5}}{\rho^2 \tau^2}\,, \qquad V_{KK} =  \frac{A_{KK}}{\rho \tau^2}\,,\quad V_{KKO} = -\frac{A_{KKO}}{\rho \tau^2}\,,
\ee
where all the $A$'s are positive. We are not planning to add any of these, however, since it was discovered recently that KK-monopoles can lead to seemingly stable dS solutions \cite{Blaback:2018hdo}, we will re-analyze our scaling limit results above in the presence of KK-monopoles. 

This task is actually trivial. The KK-monopoles or the corresponding KKO-planes scale in exactly the same way as the curvature. This means that, at the level of the overall volume and string coupling, there is no actual difference in the scalar potential, if we add them. We simply group together $A_{R_6}$, $A_{KK}$ and $A_{KKO}$ as a new effective curvature term $\bar{A}_{R_6}=A_{R_6}+A_{KK}-A_{KKO}$. Then our results from above apply and dS vacua can only exist, if the new effective curvature term is positive $\bar{A}_{R_6}>0$. Likewise, we find that the dS solutions with KK-sources that were discovered in \cite{Blaback:2018hdo} suffer from the same absence of a parametrically large volume and weak coupling limit as the ones without these sources. However, again a large number of O6-planes could lead to controlled solutions. 

Let us also discuss the inclusion of NS5-branes and the corresponding NSO5-planes. These introduce genuinely new terms in the scalar potential.\footnote{It would be interesting to check whether these can remove the obstinate tachyon or whether that is only possible by adding anti-D6-branes \cite{Kallosh:2018nrk} or KK-monopoles \cite{Blaback:2018hdo}.} If we group as above, the potential KK-sources together with the curvature, then the scalar potential takes the form\be
\label{eq:V_extended}
V(\rho,\tau) = \frac{A_H}{\rho^3 \tau^2} + \sum_{p=0,2,4,6} \frac{A_{F_p}}{\rho^{p-3} \tau^4} - \frac{A_{sources}}{\tau^3}+\frac{\bar A_{R_6}}{\rho \tau^2} + \frac{A_{NS5}}{\rho^2 \tau^2} - \frac{A_{NSO5}}{\rho^2 \tau^2}\,.
\ee
Now we minimize with respect to $\tau$
\be
\label{eq:pd}
0=\partial_\tau V = -2 \frac{A_H}{\rho^3 \tau^3} -4\sum_{p=0,2,4,6} \frac{A_{F_p}}{\rho^{p-3} \tau^5} + 3 \frac{A_{sources}}{\tau^4}-2 \frac{\bar{A}_{R_6}}{\rho \tau^3} -2 \frac{A_{NS5}}{\rho^2 \tau^3} +2 \frac{A_{NSO5}}{\rho^2 \tau^3}\,.
\ee
This is solved by
\be 
 \frac{A_H}{\rho^3 \tau^2}  + \frac{\bar{A}_{R_6}}{\rho \tau^2} +  \frac{A_{NS5}}{\rho^2 \tau^2} -  \frac{A_{NSO5}}{\rho^2 \tau^2} = -2\sum_{p=0,2,4,6} \frac{A_{F_p}}{\rho^{p-3} \tau^4} + \frac{3}{2} \frac{A_{sources}}{\tau^3}\,.
\ee
Substituting this back into the potential we find
\be
V(\rho,\tau) = -\sum_{p=0,2,4,6} \frac{A_{F_p}}{\rho^{p-3} \tau^4} + \frac{1}{2} \frac{A_{sources}}{\tau^3}\,.
\ee
We see that we again need to impose $A_{sources} \propto 2N_{O6}-N_{D6}-N_{\Db}>0$ in order to have $V>0$. This means we still need O6-planes as negative tension objects, despite having allowed for NSO5-planes with negative tension. Similarly, one can check that the inclusion of only NS5-branes still forces one to have $A_{F_0} \neq 0$, since otherwise $(-\rho\partial_\rho-\tau \partial_\tau) V\geq 3V$. However, it seems that the inclusion of net NSO5-planes, i.e $A_{NSO5}>A_{NS5}$ allows for vanishing mass parameter ($A_{F_0}=0$) in type IIA, while in principle still allowing dS vacua. This is an interesting point that deserves further study since solutions without mass parameter can be lifted and studied in M-theory. With regards to the curvature we only find the constraint that a net number of NSO5-planes, $A_{NSO5}>A_{NS5}$, is not compatible with positive or zero curvature, $A_{R_6}\propto -R_6\leq0$, since in that case $(-\rho\partial_\rho-3\tau \partial_\tau) V\geq 9V$. Keeping that in mind, the relevant terms for the scaling come from the internal curvature term, the O6-planes in the term $A_{sources}$ and the NSO5-planes. The relevant part of the potential we consider is
\be
V = \frac{A_{R_6}}{\rho \tau ^2} - \frac{A_{sources}}{\tau ^3} - \frac{A_{NSO5}}{\rho ^2 \tau ^2},
\ee
where the number of O6-planes and NSO5-planes is fixed. In the scaling limit this then gives the conditions
\be 
c_{R6} -c_{\rho} - 2 c_{\tau} = -3 c_{\tau} = - 2 c_{\rho} - 2 c_{\tau},
\ee
which yields
\be 
c_{\tau} = 2 c_{\rho} \qquad \text{and} \qquad c_{R_6} = -c_{\rho}.
\ee
In the scaling limit $\lambda \to \infty$ for $c_\rho >0$ this then implies $A_{R_6} \propto -R_6 \propto \lambda^{c_{R_6}} = \lambda^{-c_{\rho}} \rightarrow 0$. Since the curvature $R_6$ is fixed for a given manifold and cannot become parametrically small, this prohibits dS solution with parametric control. Thus a small, fixed number of NSO5-planes is not sufficient to improve our situation.\footnote{In the presence of KK-monopoles and/or KK-planes it might be possible to make the effective curvature term $\bar{A}_{R_6}=A_{R_6}+A_{KK}-A_{KKO}$ arbitrarily small. We leave a detailed study of this possibility to the future.}

There are several other ingredients that have appeared in the literature. In particular, it has long been known that the inclusion of so called non-geometric $Q$- and $R$-fluxes can lead to stable dS vacua \cite{deCarlos:2009fq, deCarlos:2009qm, Danielsson:2012by, Blaback:2013ht, Damian:2013dq, Blaback:2015zra}. These fluxes scale like
\be
V_Q = \frac{A_Q}{\rho^{-1} \tau^2}\,, \qquad \quad V_R = \frac{A_R}{\rho^{-3} \tau^2}\,.
\ee
There exist by duality also the corresponding sources for these fluxes. However, since these compactifications are non-geometric it is unclear how one can control $\alpha'$ corrections after including these fluxes or sources.

While the most commonly studied compactifications of massive type IIA are on SU(3)-structure spaces that have no 1- and 5-cycles, one can for other compactification spaces in principle also include D$p$-branes and O$p$-planes for different values of $p$. These scale as follows
\be
V_{Dp / \overline{Dp}} = \frac{A_{Dp / \overline{Dp}}}{\rho^{\frac{6-p}{2}} \tau^3}\,, \quad \qquad V_{Op} = - \frac{A_{Op}}{\rho^{\frac{6-p}{2}} \tau^3}\,.
\ee
Other ingredients that have appeared in the literature \cite{Silverstein:2007ac} and that can be supported by certain compactification spaces are fractional Wilson lines and fractional Chern-Simons invariants.

\section{An explicit example}
We would now like to complement our fully general analysis above with a simple concrete example. The example is the so called isotropic compactification of massive type IIA on $S^3\times S^3/\mathbb{Z}_2\times \mathbb{Z}_2$. This model has been studied in \cite{Villadoro:2005cu, Camara:2005dc, Aldazabal:2007sn, Flauger:2008ad, Caviezel:2008tf, Danielsson:2009ff, Danielsson:2010bc, Danielsson:2012et, Blaback:2013fca, Junghans:2016uvg, Junghans:2016abx, Roupec:2018mbn, Kallosh:2018nrk, Blaback:2018hdo} and is probably the simplest non-trivial example that gives rise to dS critical points and potentially even to stable dS vacua, if one includes anti-D6-branes or KK-monopole sources \cite{Kallosh:2018nrk, Blaback:2018hdo}. The model is the familiar `STU'-model that has three complex moduli. These arise from a compactification on $T^6/\mathbb{Z}_2\times \mathbb{Z}_2$, if we identify the three $T^2$'s in $T^6$. In our model we include metric fluxes, which effectively amounts to changing the internal space to $S^3 \times S^3$. We also turn on $H$-flux, $F_0$-flux and $F_2$-flux. The $F_4$- and $F_6$-fluxes can be set to zero in this model by shifting the moduli. The metric fluxes are set to one since the overall size is encoded in the real part of the K\"ahler modulus $T$. The complex structure moduli and the the dilaton are packaged with their axionic partners into two complex scalar fields that we call $Z_1$ and $Z_2$. These moduli are connected to the above variables $\rho$ and $\tau$ via $\rho = \text{Re}(T)$ and $\tau^4 = \text{Re}(Z_1) \text{Re}(Z_2)^3$. 

Following \cite{Kallosh:2018nrk} we will allow for anti-D6-branes that can wrap two different 3-cycles and we label their numbers by $N_{\Db,1}$ and $N_{\Db,2}$. The anti-D6-branes spontaneously break supersymmetry and their contributions cannot be packaged into the K\"ahler and the superpotential $W$, if one only uses standard superfields. For an anti-D3-brane, it was shown that one can obtain its action by using a so called nilpotent chiral multiplet $X$ that satisfies $X^2=0$ \cite{Ferrara:2014kva, Kallosh:2014wsa, Bergshoeff:2015jxa}. This result was recently extend to all anti-branes in flux compactifications down to 4d  $\N=1$ \cite{Kallosh:2018nrk}. The resulting four dimensional K\"ahler and superpotential for this model are then given by
\ba\label{eq:Kpot}
K &=& -\log\left[ \left( T + \bar{T}\right)^3 - \frac12 \frac{X \bar{X}}{N_{\Db,1} (Z_1 + \bar{Z}_1) + N_{\Db,2} (Z_2 + \bar{Z}_2)} \right] - \log\left[2^{-4} (Z_1 +\bar{Z}_1 )(Z_2 + \bar{Z}_2)^3 \right]\,,\cr
W &=& -3 f_2 T^2 + \rmi f_0 T^3 + ( \rmi h - 3 T )Z_1 -3 (\rmi h +T ) Z_2 +X\,.
\ea
From this one derives the scalar potential in the usual way, except that one has to set $X=\bar X = 0$ in the end, since $X$ only contains fermionic degrees of freedom and we are interested in the bosonic scalar potential:
\be \label{eq:scalpot}
V= e^K \left[ D_IW K^{I\bar J}\overline{D_J W} - 3 W\bar{W} \right]_{X\to 0\, , \, \bar{X} \to 0}\,.
\ee

\subsection{Rescaling symmetries of the scalar potential}
In this subsection we show that our model has two scaling freedoms, one of which corresponds to the one in the previous section, where we explored the possibility of a large number of O6-planes. In general we can scale all moduli and fluxes appearing in \eqref{eq:Kpot} as we want, without changing the presence of critical points and the eigenvalues of the mass matrix, as long as the scalar potential \eqref{eq:scalpot} only changes by an overall factor. Let us choose our two rescalings such that $T \to a T$ and $f_0 \to b f_0$ for real $a$ and $b$. These are the most general rescalings that change the superpotential only by an overall factor, which is required for consistency. In particular, this fixes all moduli to scale as
\be \label{eq:modscale}
T \to a T \,, \quad Z_1 \to a^{2} b Z_1  \,, \quad Z_2 \to a^{2} b Z_2 \,, \quad  X \to a^3 b X\,,
\ee
and likewise for their complex conjugates. Note that since Re$(T)$, Re$(Z_1)$ and Re$(Z_2)$ have to all be positive, we have to restrict to $a,b>0$. Simultaneously with the moduli we have to scale all the parameters as
\be \label{eq:fluxscale}
f_0 \to b f_0 \,, \quad f_2 \to a b f_2\,, \quad  h \to a h\,, \quad  N_{\Db,1} \to a b N_{\Db,1}\,, \quad  N_{\Db,2} \to a b N_{\Db,2} \,.
\ee
In terms of the superpotential and scalar potential this amounts to the overall rescalings
\be 
W \to a^3 b W \qquad \text{and} \qquad V \to a^{-5} b^{-2} V.
\ee
While such scalings change the physics by changing the value of the cosmological constant, i.e. the value of $V$ at its minimum, they do not change the existence and properties of critical points. So it is often simpler to fix the scaling symmetries by setting certain fluxes to unity. Then one can minimize the scalar potential and once one has found an interesting critical point, one can use the scaling symmetries to set the fluxes to any desired value.

Let us now make the link to the simple scaling example we discussed at the end of subsection \ref{ssec:dS}. There we had the flux $F_0$, represented here by the flux parameter $f_0$, and $R_6$ non-scaling. This corresponds to the special case of $b=0$ in \eqref{eq:fluxscale}. Conveniently, if we identify $a = N_{O6}$ we will quickly recover our result and show that the formulations correspond to each other. For this we need to mention that $\V_6 \sim Re(T)^3$ and $e^{-\phi} \sim \sqrt[4]{Re(Z_1) Re(Z_2)^3}/ \sqrt{\V_6}$. Using the rescalings \eqref{eq:modscale} and \eqref{eq:fluxscale} we find
\be
\V_6 \propto N_{O6}^3 \qquad \text{and} \qquad e^{-\phi} \propto \sqrt{N_{O6}},
\ee
in accordance with what we had before.

\subsection{Explicit solutions}
Having the explicit scalar potential in equation \eqref{eq:scalpot}, we find it instructive to first map out the regions of parameter space that allow for minima with positive values of $V$. In principle this is rather cumbersome even for this simple model with only six real moduli. However, there are a variety of tricks that one can use and in particular, if one adds anti-D6-branes, one can actually find the analytic solution that minimizes the scalar potential. 

First let us note that rescaling the moduli as well as the flux and brane parameters that we discussed in the previous subsection, only changes $V$ and its derivatives by an overall positive constant. Thus we can for example choose the positive parameters $a$ and $b$ such that $|f_0|=|f_2|=1$. Then we are left with a three parameter moduli space spanned by $h$, $N_{\Db,1}$ and $N_{\Db,2}$. Now we solve $\partial_{\text{Im}(T)} V = \partial_{\text{Im}(Z_1)} V = \partial_{\text{Im}(Z_2)} V=0$ in term of the axionic fields Im$(T)$, Im $(Z_1)$ and Im$(Z_2)$ and find a unique solution. The remaining equations $\partial_{\text{Re}(T)} V = \partial_{\text{Re}(Z_1)} V = \partial_{\text{Re}(Z_2)} V=0$ are hard to solve in terms of the variables. However, the addition of anti-D6-branes makes the equations $\partial_{\text{Re}(Z_1)} V = \partial_{\text{Re}(Z_2)} V=0$ linear in terms of the number of anti-D6-branes $N_{\Db,1}$ and $N_{\Db,2}$. So we can implicitly solve them in terms of $N_{\Db,1}$ and $N_{\Db,2}$. Similarly, $\partial_{\text{Re}(T)} V =0$ is quadratic in the H-flux quanta $h$ so we can solve it in terms of $h$ and find two branches of solutions. So this gives us an analytic solution for the critical points of $V$. Since we have solved the equations implicitly in terms of the parameters, this solution depends on the values of the moduli Re$(T)$, Re$(Z_1)$ and Re$(Z_2)$, which is actually a nice feature.

\begin{figure}
    \centering
        \includegraphics[width=.66\textwidth]{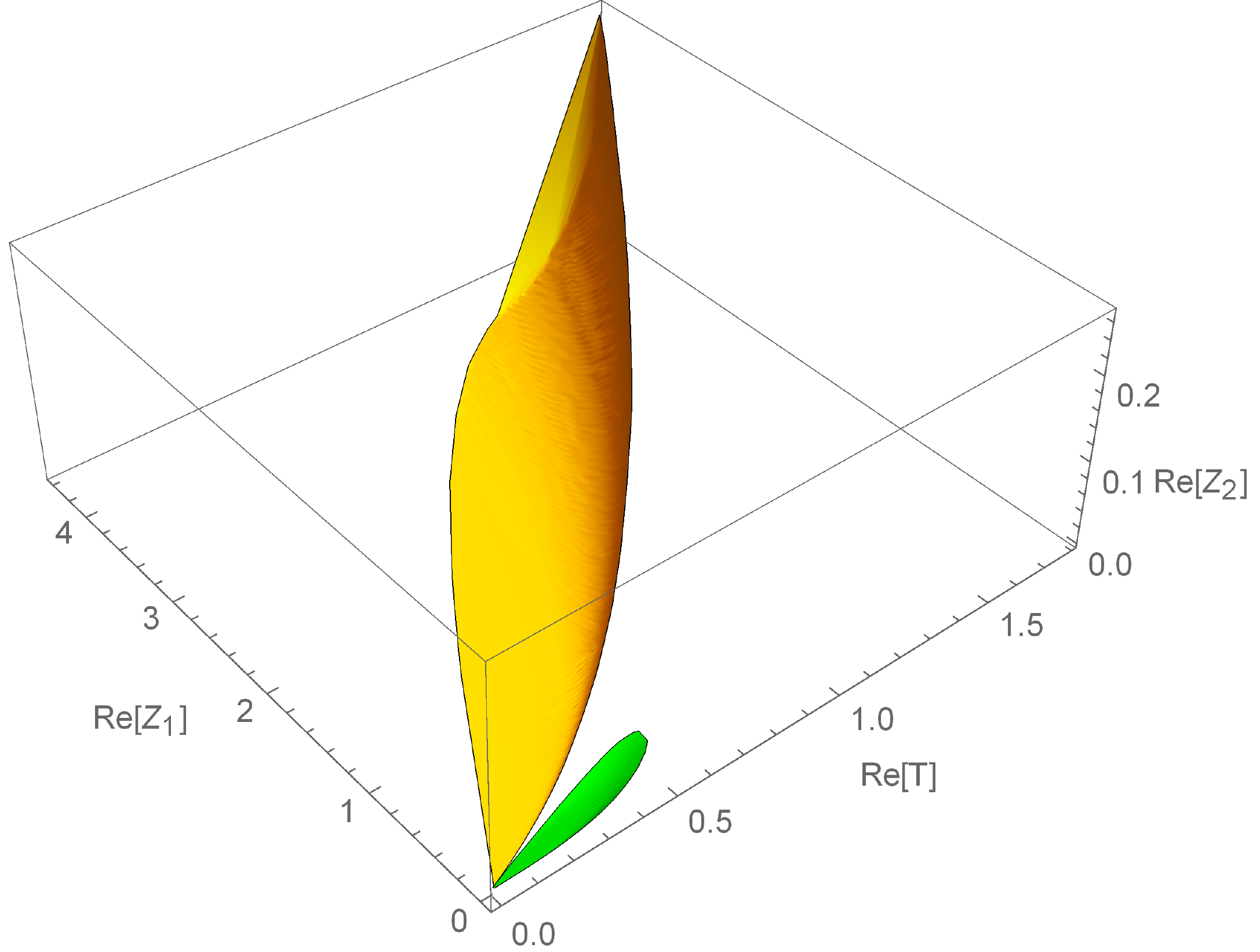}
\caption{This plot shows in yellow and green two regions for which $V$ has metastable dS minima with $N_{\Db,1}>0$ and $N_{\Db,2}>0$. The smaller green region has no tachyons even if we open up the non-isotropic directions, while the yellow regions has tachyonic directions in that case.}
  \label{fig:dsregions}
\end{figure}

We are particularly interested in solutions with $V_{\rm min}>0$ for which the six real fields all have a positive mass. We scanned through different values for the real parts of the moduli and searched for such regions. We additionally imposed that $N_{\Db,1}>0$ and $N_{\Db,2}>0$, since otherwise we would effectively have included anti-O6-planes. We have at this stage not worried about quantization of the parameters. The result of our scans are two three dimensional regions that are rather thin and that are shown in figure \ref{fig:dsregions}. The smaller green region has even no tachyonic directions, if we open up the non-isotropic directions and include all fourteen real moduli.

For any of these solutions we can use the scaling symmetries discussed in the previous subsection to change the values of the fluxes and moduli. However, the tadpole condition and the fact that $N_{O6}=1$ in this model (cf. footnote \ref{fn:NO6}), prevent us from substantially changing the values of the moduli. Thus we see from figure \ref{fig:dsregions} that there are no solutions at $\rho=$Re$(T)\gg1$ or $\tau^4 =$Re$(Z_1)$Re$(Z_2)^3\gg 1$. This means, consistent with our expectation from the previous section, that this model has no dS solutions in a trustworthy regime.

Having this explicit model and analytic control over the solutions, allows us to go beyond the general scaling we derived for the volume and the string coupling with respect to the number of O6-planes in equation \eqref{eq:largeNO6}. We can actually check what are the order one factors and whether they make it easier or harder to find solutions with large volume and weak coupling. In particular, for all the stable dS vacua shown in figure \ref{fig:dsregions} we find that the volume and the string coupling scale as follows with the number of O6-planes
\be
\V_6 = c_1 N_{O6}^{3} \qquad\quad \text{with}\qquad 8 \times 10^{-8} \lesssim c_1 \lesssim .24\,,
\ee
and 
\be
e^{-\phi} = c_2 \sqrt{N_{O6}} \qquad \text{with}\qquad 4\times 10^{-5} \lesssim c_2 \lesssim .12\,.
\ee
So the order one factors are not helping in obtaining trustworthy solutions but actually make it harder.

Let us again stress that in this particular setup the number of O6-planes is fixed to $N_{O6}=1$, so that it is not a free parameters. However, the result above should provide a reasonable estimate for the number of O6-planes one needs in these compactifications to be able to suppress $\alpha'$ and string loop corrections.

\section{Conclusion}\label{sec:conclusion}
Given the recent interest in dS vacua and dark energy in string theory we have studied the probably simplest flux compactifications that bear the hope of giving rise to stable, trustworthy dS vacua, namely we have studied compactifications of massive type IIA in the presence of fluxes and a variety of other classical ingredients. This class of models might provide the simplest setting for improving our understand of dS space as well as the recently proposed dS swampland conjecture \cite{Obied:2018sgi}. We have addressed the important question of when solutions of the scalar potential can arise at large volume and weak coupling, i.e. of when the solutions are trustworthy and receive negligible $\alpha'$ and string loop corrections. We have shown that for dS vacua one cannot get parametric (i.e. arbitrarily good) control over corrections in these compactifications. On the one hand, this might not come as a surprise since there are no constructions of dS vacua that claim to have such a parametric control and there are reasons going back to Dine-Seiberg \cite{Dine:1985he}, why one should never expect it. On the other hand, there is no mathematical proof that such solutions cannot exist (see for example the recent work \cite{Hebecker:2018vxz, Junghans:2018gdb}) and AdS vacua with parametric control have been found in compactifications of massive type IIA \cite{DeWolfe:2005uu}. In our analysis we have allowed for all RR- and NSNS-fluxes, included curvature, O6-planes and (anti-)D6-branes as well as KK-monopoles and KKO-planes. A natural extension would be to include NS5-branes and NSO5-planes. We have shown that the latter potentially allow for dS solutions without mass parameter, $F_0$. This is very interesting since such dS solutions, if they actually exist, could be lifted to M-theory.

While we cannot get arbitrary good controlled over the dS solutions, we have shown that a large number, $N_{O6}$, of O6-planes can lead to a large volume $\V_6$ and a weak string coupling:
\be\label{eq:scaling}
\V_6 \propto N_{O6}^{3} \qquad \text{and} \qquad e^{-\phi} \propto \sqrt{N_{O6}}\,.
\ee
The number of O6-planes in most of the models studied so far seems to be $N_{O6}=1$, which explains why so far no dS solutions have been found at large volume and weak coupling. It would be very interesting to find either manifolds that allow for orientifold projections with a large number of fixed points or to show that these do not exist.

We have complemented our general analysis by the discussing of a simple concrete example, namely the compactification on the isotropic $S^3\times S^3/\mathbb{Z}_2\times\mathbb{Z}_2$ orbifold. Isotropic means here that we set three of the K\"ahler moduli and three of the complex structure moduli equal so that we are only left with three complex moduli. It was recently shown that the inclusion of anti-D6-branes in the scalar potential seems to lead to metastable dS solutions \cite{Kallosh:2018nrk}. We have mapped out the three dimensional parameter space of these metastable dS solutions. We find that it consist of two pieces, one of which has no tachyons even if we open up the directions that disappeared in the isotropic truncation, while the other one has tachyons in the non-isotropic directions.

To connect to the first part of the paper, we have identified in this concrete model a scaling symmetry that changes the number of O6-planes. For this model one would have to fix this symmetry such that $N_{O6}=1$, but if one does not do that then one recovers the scaling from equation \eqref{eq:scaling} above. Since we have an explicit model we can even determine the prefactor range for the solutions. For this model we have
\be
\V_6 = c_1 N_{O6}^{3} \qquad \quad\text{with}\qquad c_1 \lesssim \frac14\,,
\ee
and 
\be
e^{-\phi} = c_2 \sqrt{N_{O6}} \qquad \text{with} \qquad  c_2 \lesssim \frac18\,.
\ee
We leave it to the future to check whether actual compactification spaces with large numbers of O6-planes give rise to dS vacua that agree with the above result and to check whether there are actually compactification spaces that give rise to dS vacua at large volume and weak string coupling.

\acknowledgments
We are grateful to D. Andriot, D. Junghans and T. Van Riet for enlightening discussions. This work is supported by FWF grants with the number P 30265 and P 28552.

\bibliographystyle{JHEP}
\bibliography{refs}

\end{document}